\documentclass[twocolumn,twocolappendix]{aastex63}

\usepackage[T1]{fontenc}
\usepackage{ae,aecompl}

\usepackage{natbib}
\usepackage{multirow}
\usepackage{diagbox}

\usepackage{footnote}
\usepackage[flushleft]{threeparttable} 
\usepackage{rotating}

\usepackage{amsmath}
\usepackage{upgreek}

\newcommand\mc[1]{\multicolumn{1}{c}{#1}}

\shorttitle{Jet production efficiency in \textit{Swift}/BAT AGNs}
\shortauthors{Rusinek et al.}

\begin{document}

\title{On the diversity of the jet production efficiency in \textit{Swift}/BAT AGNs}

\correspondingauthor{Katarzyna Rusinek}
\email{krusinek@camk.edu.pl}

\author[0000-0002-6424-6558]{Katarzyna Rusinek}
\affiliation{Nicolaus Copernicus Astronomical Center, Polish Academy of Sciences, Bartycka 18, 00-716 Warsaw, Poland}

\author{Marek Sikora}
\affiliation{Nicolaus Copernicus Astronomical Center, Polish Academy of Sciences, Bartycka 18, 00-716 Warsaw, Poland}

\author[0000-0003-4323-0984]{Dorota Kozie{\l}-Wierzbowska}
\affiliation{Astronomical Observatory, Jagiellonian University, Orla 171, 30-244 Krak\'{o}w, Poland}

\author{Maitrayee Gupta}
\affiliation{Nicolaus Copernicus Astronomical Center, Polish Academy of Sciences, Bartycka 18, 00-716 Warsaw, Poland}
\affiliation{Harvard Smithsonian Center for Astrophysics, 60 Garden Street, Cambridge, MA 02138, USA}


\begin{abstract}

We address the very large diversity of the jet production efficiency in active galactic nuclei (AGNs) by using data on low redshift AGNs selected from the \textit{Swift}/BAT catalog and having black hole (BH) masses larger than $10^{8.5}\,M_{\odot}$. Most of these AGNs accrete at intermediate rates and have bolometric luminosities dominated by mid-IR radiation. Our sample contains $14\%$ radio-loud (RL), $6\%$ radio-intermediate (RI), and $80\%$ radio-quiet (RQ) AGNs. All RL objects are found to have extended radio structures and most of them have classical FR\,II morphology. Converting their radio loudness to the jet production efficiency, we find that the median of this efficiency is on the order of $(\epsilon_d/0.1)\%$, where $\epsilon_d=L_{\rm bol}/\dot{M}c^2$ is the radiation efficiency of the accretion disk. Without knowing the contribution of jets to the radio emission in the RQ AGNs, we are only able to estimate their efficiencies using upper limits. Their median is found to be $0.002(\epsilon_d/0.1)\%$. Our results suggest that some threshold conditions must be satisfied to allow production of strong, relativistic jets in RL AGNs. We discuss several possible scenarios and argue that the production of collimated, relativistic jets must involve the Blandford-Znajek mechanism and can be activated only in those AGNs whose lifetime is longer than the time required to enter the magnetically arrested disk (MAD). Presuming that MAD is required to collimate relativistic jets, we expect that the weak nonrelativistic jets observed in some RQ AGNs are produced by accretion disks rather than by rotating BHs.

\end{abstract}


\section{Introduction}

While the first quasars were discovered following the identification of some radio sources with optical point sources located at cosmological distances \citep{Schmidt1963}, it quickly turned out that most of them were radio-quiet \citep{Sandage1965}. This led to their division into radio-loud quasars (RLQ) and radio-quiet quasars \citep[RQQ,][]{Kellermann1989} with an aproximate number proportion 1:10. However, later studies using deeper radio surveys led to the discovery of many quasars with intermediate radio loudness and the often-claimed radio bimodality came into question \citep[see][ and refs. therein]{Kratzer2015}.

A broad distribution of radio loudness was also found in active galactic nuclei (AGNs) located at much closer distances than luminous quasars \citep[e.g.][]{Rafter2009}, and if the dominating radio flux of the extended radio sources with which they were associated was included, the bimodality reappeared \citep{Rafter2011}. A bimodal radio distribution in AGNs was also confirmed by the recent studies of \citet{Gupta2018,Gupta2020}. In order to avoid biases in the determination of the radio loudness distribution associated with optical and radio selection limits, \citeauthor{Gupta2018} based their studies on using a sample of AGNs selected from the \textit{Swift}/BAT catalog \citep{Ricci2017}. Due to the very low sensitivity of the BAT detector, most of these AGNs are located at low redshift. The radio-loudest AGNs were found, like in quasars, in AGNs with $M_{\rm BH} > 10^8 M_{\odot}$, but with much lower accretion rates. The latter concerns also RQ AGNs with very massive black holes (BHs) and can be explained by the  ``downsizing effect", according to which the average specific accretion rates in massive AGNs decrease  with decreasing redshift \citep[e.g.][]{Fanidakis2012}. In order to minimize the impact of   RL and RQ AGNs having different average black hole masses and Eddington ratios ($\lambda_{\rm Edd} = L_{\rm bol}/L_{\rm Edd}$) on the compared radio properties, the sample adopted from the \textit{Swift}/BAT catalog of AGNs was reduced to have RQ and RL AGNs with similar ranges of these parameters. 

The advantages of studying the radio properties of AGNs and possible relations to the optical properties of their host galaxies and environments using the low redshift samples are obvious: (1) almost all of these AGNs have radio detections; (2) studies of radio morphology are not limited to only extended radio sources; (3) studies of optical morphology of their hosts and environment are possible; (4) biases associated with cosmological evolution are minimized; (5) a much lower probability of having radio-intermediate (RI) AGNs dominated by strong starbursts and accretion disk winds. Hence, studies incorporating such samples have an exceptional potential to provide a variety of constrains which can be used to select the  most promising scenario to explain the origin of the large diversity of the jet production efficiency.

Our work is organized as follows: in \S2 the sample is defined; in \S3 the radio loudness distribution is derived and radio morphologies are determined; in \S4 bolometric luminosities, black hole masses, Eddington ratios, and jet powers are derived and used to construct the distribution of jet production efficiency; in \S5 properties of the host galaxies are reviewed. A comparison of our results with others and their theoretical implications and possible interpretations are provided in \S6 and summarized in \S7.

Throughout the paper we assume a $\Lambda$CDM cosmology with $H_0 = 70 \ {\rm km} \ {\rm s}^{-1} \ {\rm Mpc}^{-1}$, $\Omega_m=0.3$, and $\Omega_{\Lambda}=0.70$.

\section{The sample}
\label{sec_SAMPLE}

Our initial sample is taken from \citet{Gupta2018, Gupta2020} who performed a comparison of RL and RQ, both Type 1 and Type 2, AGNs in various spectral bands. The sources in their sample were selected from the BAT AGN Spectroscopic Survey \citep[BASS,][]{Ricci2017}, by excluding blazars, Compton-thick ($\log N_{\rm H} > 24$) AGNs, and sources with missing optical spectroscopic classification in \citet{Koss2017} which resulted in 664 objects. The sample in \citet{Gupta2018} was limited to 70 sources. This was due to the limits imposed on BH masses and Eddington ratios ($8.48 \leq \log M_{\rm BH} \leq 9.5$, $-2.55 \leq \log \lambda_{\rm Edd} \leq -1$) which gave the authors possibly the best representation of radio galaxies and their radio-quiet counterparts working at intermediate accretion rates. In \citet{Gupta2020}, some of the earlier restrictions were relaxed and additionally, using the relation between BH masses and the NIR luminosities of host galaxies of AGNs \citep{MarconiHunt2003,Graham2007} the BH masses were calculated. As a result, instead of studying only those sources for which the BH masses are known the sample studied in \citet{Gupta2020} has 290 objects, all of them with $M_{\rm BH} \geq 10^{8.5} M_{\odot}$, $z \leq 0.35$ and $0.001 \leq \lambda_{\rm Edd} \leq 0.03$. Hence, we decided to make use of the sample from \citet{Gupta2020} and complete it by adding radio-intermediate AGNs \citep[which were excluded by the authors; see][]{Gupta2018,Gupta2020} choosing them in the same way as the other objects were found. This resulted in finding 24 additional sources (out of which 4 turned out to be RQ) giving us a final sample consisting of 314 \textit{Swift}/BAT AGNs.

The detailed description of the data and the procedure used to build this sample, originally described in \citet{Gupta2018,Gupta2020}, is reiterated in Appendix \ref{appendix_DATA}.

\section{Radio properties} 
\label{sec_RADIO}

\subsection{Radio Loudness}
\label{subsec_RADIO_radio_loudness}

Our radio loudness parameter, as in \citet{Gupta2020}, is given as $R = F_{1.4}/F_{\nu_{\rm W3}}$, where $F_{1.4}$ and $F_{\nu_{\rm W3}}$ are the monochromatic fluxes at 1.4\,GHz\footnote{Those are taken from NVSS, FIRST and SUMSS, where radio fluxes from the latest, at 843\,MHz, were recalibrated to 1.4\,GHz using a radio spectral index of $\alpha_r = 0.8$ and the convention of $F_{\nu} \propto \nu^{-\alpha}$; see Appendix \ref{subsec_appendix_DATA_radio}.} and in WISE at $\nu_{\rm W3} = 2.5 \times 10^{13}$\,Hz, respectively. This relates to the definition given by \citet{Kellermann1989}, i.e. $R_{\rm KL} = F_{5}/F_{\nu_{\rm B}}$, where $F_{5}$ and $F_{\nu_{\rm B}}$ are the monochromatic fluxes at 5\,GHz and in B band ($6.8 \times 10^{14}$\,Hz), as $R \approx 0.1 \times R_{\rm KL}$, assuming the spectral indices of $\alpha_{1.4-5} = 0.8$ and $\alpha_{\nu_{\rm B} - \nu_{\rm W3}} = 1$ \citep[see][]{Gupta2018}. Based on this we partitioned the sources into the following  radio classes, strictly corresponding to those in \citet{Kellermann1989}: radio-loud when $R > 10$; radio-intermediate for $1 < R < 10$; radio-quiet when $R < 1$. 

\begin{figure}[t]
\begin{center}
\includegraphics[width=0.45\textwidth]{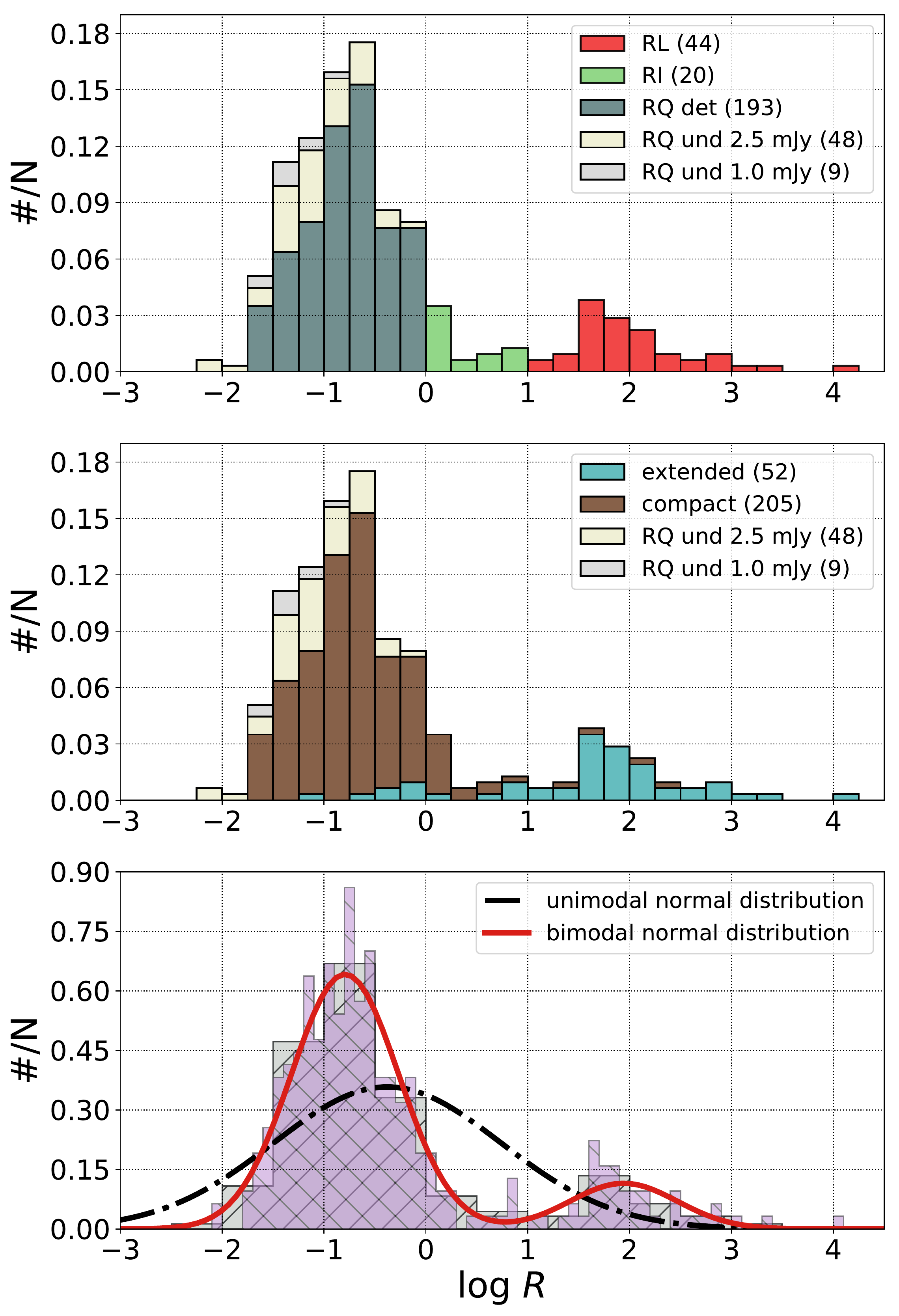}
\caption{The radio loudness distribution for \textit{Swift}/BAT AGNs in our sample. The top panel shows groups of radio-loud (red), radio-intermediate (green) and radio-quiet (grey) sources, based on the definition of the radio loudness parameter given as $R = F_{1.4}/F_{\nu_{\rm W3}}$, which, for given classes, takes values of $R >10$, $1 < R < 10$ and $R<1$, with its median value of $\log R$ equal to $1.88$, $0.23$ and $-0.84$, respectively. In RQ class we separate sources with (RQ detected, dark grey, with median value of $\log R \simeq -0.75$) and without radio detections, dividing the latter group into two categories: RQ undetected with a 2.5\,mJy detection upper limit (light yellow), when the source is in the footprint of NVSS or SUMSS; and RQ undetected up to 1.\,mJy (light grey), when the source is in the FIRST area. The middle panel distinguishes between sources with (teal) and without (brown) extended radio emission together with the same groups of RQ undetected AGNs as mentioned earlier. Here the median values of $\log R$ are $1.74$ and $-0.72$ for extended and compact AGNs, respectively. The characteristics of these radio morphologies is closely described in Sections \ref{subsubsec_RADIO_radio_morphologies_compact} and \ref{subsubsec_RADIO_radio_morphologies_extended} and exact numbers are presented in Table~\ref{tbl_1_radio_morph}. The bottom panel shows the sample distribution using two different bin widths in grey and purple.  Overplotted are maximum likelihood estimates for unimodal and bimodal normal distributions in black and red, respectively.}
\label{fig_1_hist_R}
\end{center}
\end{figure}

Since we used three radio catalogs differing in radio wavelength, angular resolution, and sensitivity, we decided to adopt the radio flux from NVSS, whose data is available for most of the sources and whose sensivity accounts for all of the extended emission. For objects lacking NVSS data we took fluxes from SUMSS and FIRST\footnote{The exception being the group of 57 compact sources, for which, even though they have data in both, NVSS and FIRST, we decided to take FIRST fluxes. The explanation is given in Appendix~\ref{appendix_SIZE}. This choice however does not affect the ascription to their radio classes.}. Such an approach resulted in finding 44 RL, 20 RI and 250 RQ AGNs in our sample, with their radio loudness medians of $75.40$, $1.69$ and $0.15$, respectively. The exact radio loudness distribution is presented in the top panel of Fig.~\ref{fig_1_hist_R}. 

Among 314 \textit{Swift}/BAT AGNs, 257 of them have and 57 lack radio data. As one can see in Fig.~\ref{fig_1_hist_R}, those radio undetected sources belong entirely to the RQ class. In the group of radio detected objects we can specify two subsamples of sources for which we have both: (1) NVSS and SUMSS data (12 sources); (2) NVSS and FIRST data (76 sources). While the NVSS and SUMSS total fluxes (compared at 1.4\,GHz) are almost the same, the ratio of NVSS to FIRST total fluxes is slightly more significant, with a median of 1.2, showing that indeed, some of the faint extended radio emission might be lost while using only higher angular resolution and better sensitivity radio data.

\subsection{Radio Morphologies}
\label{subsec_RADIO_radio_morphologies}

Within the group of 257 radio detected AGNs we can distinguish two main subsamples -- those with and without extended radio emission, represented by 52 and 205 objects, respectively. Below we give detailed characteristics of our radio morphological classification.

\subsubsection{Compact Sources}
\label{subsubsec_RADIO_radio_morphologies_compact}

Sources belonging to the group of compact AGNs are defined as those for which only one radio match, with its location corresponding to the optical center, was found. Based on whether the accurate size of the fitted major axis after deconvolution in a given radio catalog was available or not, compact sources form two groups: \textit{resolved} and \textit{unresolved}, consisting of 91 and 114 sources, respectively.

\subsubsection{Extended Sources}
\label{subsubsec_RADIO_radio_morphologies_extended}

All AGNs with more than one confirmed radio match are classified as extended. Based on the appearance of their radio morphologies, this subsample has been divided into the following morphological groups:
\begin{enumerate}
\item \textit{complex}, in which we include sources with multiple pairs of lobes, i.e.~double-double radio galaxies (DDRG) and X-shaped sources (5 sources)\footnote{Those are: PKS 0707-35; B2 1204+34; 3C 403; 3C 445; PKS 2356-61.},
\item \textit{triple}, when the core and a pair of lobes are clearly visible (20 sources),
\item \textit{double}, objects with a pair of lobes but with no detection of a core corresponding to the optical centre (15 sources),
\item \textit{knotty}, sources with quite extended, yet difficult to define, emission present on the radio map (12 sources).
\end{enumerate}

In general the extended radio emission from AGNs in our sample can be distinguished into sources with (complex, triple, double) and without (knotty) visible radio lobes.

\subsubsection{Radio Morphology vs. Radio Loudness Classification}
\label{subsubsec_RADIO_radio_morphologies_comparison}

Using the above described division of radio detected AGNs, we checked how our radio morphological groups relate to the radio loudness categorization. In Table \ref{tbl_1_radio_morph} we list this characteristics. Most of the RL sources are found to have lobed radio morphologies (36 out of 44)\footnote{From the literature we found that 4 out of 8 RL AGNs without lobes have, in fact, double radio structures. Those objects are: PKS 0222-23 in \citet{Kapahi1998b}; PKS 0326-288 in \citet{Kapahi199a}; [HB89] 1130+106 in \citet{Nilsson1998}; PKS 1916-300 in \citet{Duncan1992}.}, while almost all RQ objects correspond to compact sources (186 out of 193). In the case of RI objects, the ratio of sources with extended radio emission to those without is 1:3. This shows that the fraction of AGNs with extended radio emission decreases along with their radio loudness, which is not only clearly visible in Fig.~\ref{fig_1_hist_R} but is also reflected by the quite similar median values of the radio loudness of RL and extended sources ($75.40$ and $54.61$) and of RQ detected and compact sources ($0.18$ and $0.19$). 

In addition to the above we analyzed whether a unimodal or bimodal normal distribution best corresponds to our radio loudness sample. We performed a maximum likelihood estimate to fit the data to the two distributions, and then conducted a Kolmogorov-Smirnov test with p-values of $3.13 \times 10^{-10}$ and $0.87$ for the uni- and bimodal distributions, respectively. We found that while the unimodal distribution differs significantly from the observed sample, the same cannot be said about the bimodal distribution. The fitted distributions are shown on the bottom panel on Fig.~\ref{fig_1_hist_R}.

\subsubsection{FR I/II Classification}
\label{subsubsec_RADIO_radio_morphologies_FR}

For sources with lobed radio morphology (i.e.~complex, triple, and double objects; 40 AGNs total) based on the appearance of their radio maps, we established their Fanaroff-Riley classification \citep{FanaroffRiley1974} finding the following: 4 objects of type FR\,I; 2 AGNs of mixed FR\,I/II class; and 34 type FR\,II AGNs. Our classification is in an agreement with the data given in e.g. \citet{Rafter2011}, \citet{Koziel-Wierzbowska2011} and \citet{Panessa2016}.

\subsection{Physical Sizes}
\label{subsec_RADIO_radio_morphologies_size}

\begin{table}[t]
\begin{center}
\caption{Radio morphologies and radio classes of radio detected sources (257 objects) in our sample of \textit{Swift}/BAT AGNs. A detailed description is given in Section~\ref{subsec_RADIO_radio_morphologies}.}
\label{tbl_1_radio_morph}
\begin{tabular}{lrcccc} \hline
\multicolumn{2}{c}{\multirow{2}{*}{Radio Morphology}} & \multicolumn{3}{c}{Radio Class} & \multirow{2}{*}{Total} \\ \cline{3-5}
\multicolumn{2}{c}{}                   & RL & RI & RQ  &      \\ \hline \hline
\multirow{4}{*}{Extended} & Complex    &  5 &    &     &   5  \\ 
                          & Triple     & 16 &  4 &     &  20  \\ 
                          & Double     & 15 &    &     &  15  \\ 
                          & Knotty     &  4 &  1 &   7 &  12  \\ \hline
\multirow{2}{*}{Compact}  & Resolved   &  4 &  3 &  84 &  91  \\
                          & Unresolved &    & 12 & 102 & 114  \\ \hline \hline
\multicolumn{2}{c}{Total}              & 44 & 20 & 193 & 257  \\ \hline
\end{tabular}
\end{center}
\end{table}

The (projected) largest linear size (LLS) was determined for each radio detected source in our sample. Noting that the true definition of LLS is given as the distance between two hotspots, the most reliable calculations are available for sources with double radio lobes. For complex AGNs with multiple pairs of lobes, the pair with the largest separation was used. The sizes of knotty radio sources correspond either to the distances between the most distant radio matches, or to our measurements obtained from their radio maps. The sizes of compact AGNs (exact measurements for resolved and upper limits for unresolved objects) were taken directly from the radio catalogs as the major axis of the fitted Gaussian (a detailed procedure for the size estimation of compact AGNs is given in Appendix~\ref{appendix_SIZE}). 

\begin{figure*}[ht!]
\begin{center}
\includegraphics[width=0.98\textwidth]{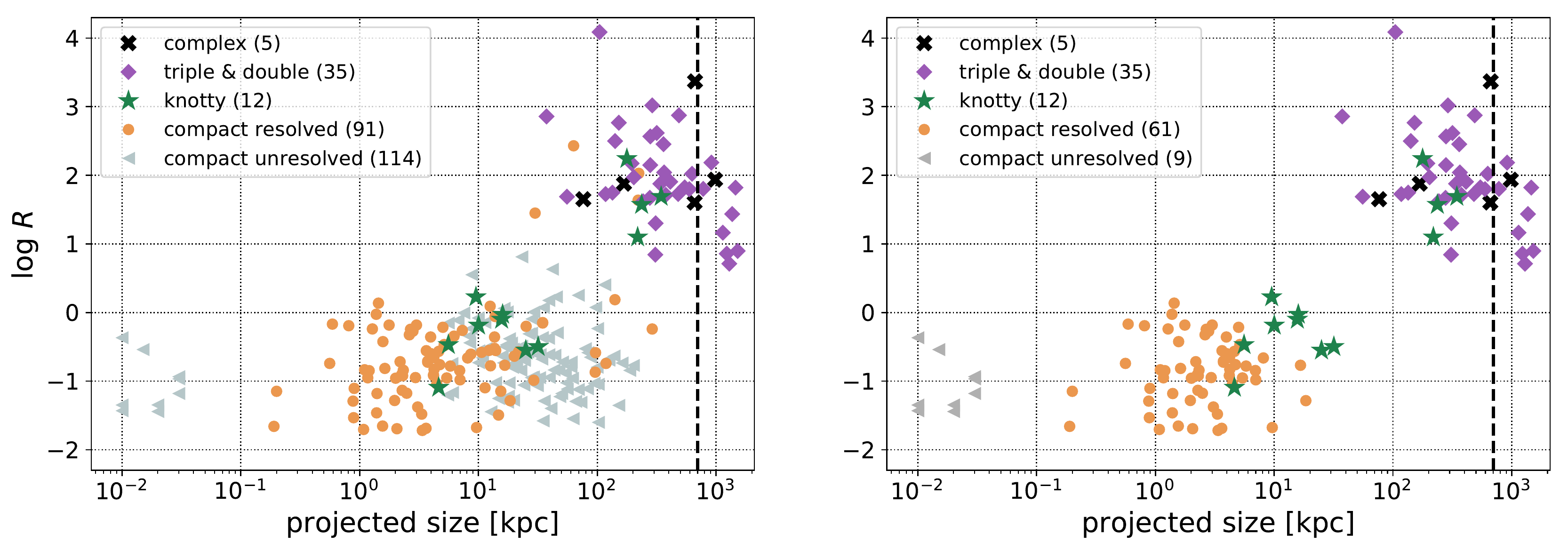}
\caption{The distribution of radio detected sources in the $\log R$ vs. projected size plane. Morphological groups are presented as follows: complex correspond to black crosses; triples and doubles are presented together as purple diamonds; knotty objects are marked as green stars; the group of compact resolved sources is shown as orange circles while compact unresolved ones correspond to grey triangles. The left panel presents all radio detected objects (257 AGNs). Compact sources on the right panel have been limited to those with FIRST data used (122 AGNs total) which allows a clear separation to be seen between resolved and unresolved objects. Nine of our sources have sizes corresponding to giant radio galaxies (GRGs) and those are located to the right of the vertical dashed line marking the size of 700\,Mpc.}
\label{fig_2_size_radio}
\end{center}
\end{figure*}

The results for the largest linear size calculations for all the radio detected sources are shown in the left panel of Fig.~\ref{fig_2_size_radio}. Knowing that only 3 out of 6 morphological groups (namely: complex, triple and double sources) presented in this paper have direct measurements of projected size, and the sizes of compact (unresolved and resolved)  and knotty sources are upper limits (i.e. are possibly shifted to the left on this chart), a clear trend confirming our results from Section~\ref{subsubsec_RADIO_radio_morphologies_comparison} is found, namely -- the most extended AGNs are radio-loud whereas the radio-quiet ones do not achieve such big sizes. This distinction is especially visible in the right panel of Fig.~\ref{fig_2_size_radio} where only compact sources with radio data from FIRST (which, within the group of 57 compact sources, provides a more precise, and on average 15 times smaller sizes than NVSS; see Appendix~\ref{appendix_SIZE}) are included. This figure shows that compact resolved and compact unresolved objects are in fact located in different parts of the diagram.

From Fig.~\ref{fig_2_size_radio} we note that 9 sources have projected sizes beyond 700\,kpc, which is the definition of giant radio galaxies (GRGs). The fraction of GRGs in our sample compared to the total number of lobed sources ($9/40 \approx 23\%$), is almost the same as in \citet{Bassani2016}, where the authors found that 14 out of their 64 confirmed radio galaxies selected in the soft gamma-ray band and having double lobe morphologies have giant sizes. Furthermore, \citet{Bruni2019} found that 61\% of \citeauthor{Bassani2016} giant radio sources have Gigahertz-Peaked Spectrum (GPS) cores, i.e. young nuclei. We checked that all of our GRGs are in common with those studied by \citet{Bruni2019} with 6 of them ($67\%$) having GPS cores suggesting the ongoing accretion and reactivation of the jets\footnote{Those are: 2MASX J03181899+6829322; IGR 14488-4008; Mrk 1498; 4C +34.47; 4C +74.26; PKS 2356-61.}. Reversely, four of their AGNs are not found in our sample as they were not observed by \textit{Swift}/BAT or have blazar-like nuclei, while another 2 sources (namely 4C +63.22 and PKS 2356-61) listed in \citeauthor{Bruni2019} are not recognized as GRGs since their sizes, estimated in this work, are slightly below 700\,kpc.

\section{Jet production efficiency}
\label{sec_JET}

\subsection{Bolometric Luminosity}
\label{subsec_JET_bolometric_luminosity}

Our calculations of bolometric luminosity started by checking whether or not the method in which MIR W3 fluxes were used to estimate $L_{\rm bol}$, as was done in \citet{Gupta2018}, can be successfully applied to our sample of AGNs accreting at moderate accretion rate. Such a verification was possible due to the accessibility of the multi-band spectra, and in turn, the exact values of bolometric luminosities, which are presented in \citet{Gupta2020}, upon which we built our sample of \textit{Swift}/BAT AGNs. 

We decided to analyze a subsample consisting of 131 (20 RL, 111 RQ) sources -- all of those being Type 1 and having strict MIR, NIR, optical-UV, and hard X-rays detections \citep[for more details see Section 4 in ][]{Gupta2020}. This choice was dictated by our need of having possibly the most accurate estimations of $L_{\rm bol}$ in which the obscuration by the dusty torus is minimized. 

\begin{figure}[h]
\begin{center}
\includegraphics[width=0.48\textwidth]{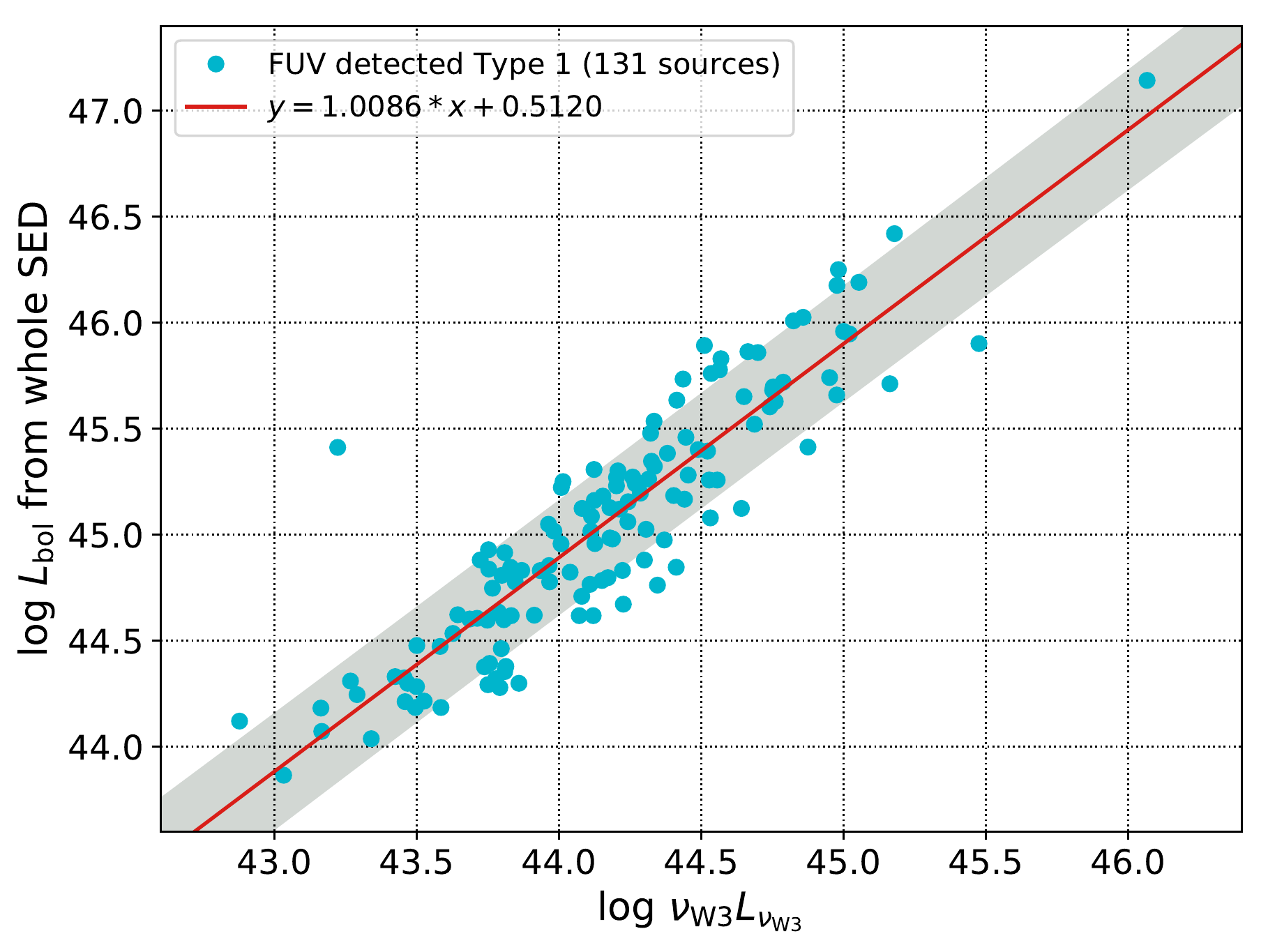}
\caption{Comparison of monochromatic luminosity in the W3 band, $\nu_{\rm W3} L_{\nu_{\rm W3}}$, and bolometric luminosity estimated from the whole SED, $L_{\rm bol}$, for 131 Type 1 AGNs, for which exact measurements (not upper limits) in FUV band were given in \citet{Gupta2020}. The red line corresponds to the linear dependence found between these two parameters via linear regression and the shaded area illustrates a $1\sigma$ confidence interval. Based on such a relation, the bolometric luminosity for all objects in our sample was calculated using Equation~(\ref{eq:Lbol_approx}).}
\label{fig_3_Lbol_estimation}
\end{center}
\end{figure}

By representing the given subsample in the $\log\,( \nu_{\rm W3} \, L_{\nu_{\rm W3}})$ vs. $\log L_{\rm bol}$ plane, where $L_{\rm bol}$ is taken from \citet[][the specific formulas can be found in Appendix B therein]{Gupta2020} and for which the distribution is shown in Fig.~\ref{fig_3_Lbol_estimation}, we found that indeed almost all sources exhibit linear correlation ($r \approx 0.90$, where $r$ is the correlation coefficient) between monochromatic luminosity in W3 band and bolometric luminosity in their logarithmic quantities\footnote{The only  source not clearly following this trend is LEDA 100168 ($\log \nu_{\rm W3} L_{\nu_{\rm W3}} = 43.22$, $\log L_{\rm bol} = 45.41$).}. Linear regression produces a formula of such a dependence which is given as
\begin{equation}
\label{eq:Lbol_exact}
L_{\rm bol}\,{\rm [erg\,s^{-1}]} = 3.25 \times \big( \nu_{\rm W3} \, L_{\nu_{\rm W3}}\,{\rm [erg\,s^{-1}]} \big) ^{1.0086},
\end{equation} 
which for $\nu_{\rm W3} \, L_{\nu_{\rm W3}}$ in the range $10^{42} - 10^{46} \, \rm {erg\,s^{-1}}$ is well approximated by
\begin{equation}
\label{eq:Lbol_approx}
L_{\rm bol}\,{\rm [erg\,s^{-1}]} \simeq 7.77 \times \nu_{\rm W3} \, L_{\nu_{\rm W3}}\,{\rm [erg\,s^{-1}]}.
\end{equation}
The median ratio of bolometric luminosity obtained from the whole SED to that calculated from Equation~\ref{eq:Lbol_approx} for a studied subsample of 131 sources is $0.96$. Hence, we conclude that the estimation of bolometric luminosity for AGNs accreting at moderate accretion rate from the monochromatic MIR luminosity is reliable and can be effectively used for our whole sample.

\subsection{Black Hole Mass and Eddington Ratio}
\label{subsec_JET_mbh_eddington_ratio}

In order to obtain BH masses in a uniform way for all AGNs in our sample, we decided to use the relation between black hole masses and near-infrared luminosities of the host galaxies \citep{MarconiHunt2003}, with the NIR data being taken from the Two Micron All Sky Survey \citep[2MASS;][]{Skrutskie2006}. The formula used for our calculations is given as $\log(M_{\rm BH}/M_{\odot}) = -0.37 \times M_{\rm K} -0.59$ \citep{Graham2007}, where $M_{\rm K}$ is the absolute K-band magnitude of the galaxy. A more specific explanation of this strategy is given in \citet[][see Appendix A therein]{Gupta2020}.

Having calculated bolometric luminosities and black hole masses, we obtained the Eddington ratio as $\lambda_{\rm Edd} = L_{\rm bol}/L_{\rm Edd}$, where $L_{\rm Edd}$ is the Eddington luminosity. Its values for the whole sample range from $0.0004 \leq \lambda_{\rm Edd} \leq 0.1905$\footnote{The difference between the ranges of $\lambda_{\rm Edd}$ in our and \citet{Gupta2020} studies results from different methods of $L_{\rm bol}$ estimation.} with a median value of $\lambda_{\rm Edd} = 0.011$ and its distribution is presented in the top histogram of Fig.~\ref{fig_4_PjLbol_Eddratio}.

\subsection{Jet Power}
\label{subsec_JET_jet_power}

Among various methods of estimating jet powers of AGNs, we decided to use the one based on the calorimetry of radio lobes, originally formulated by \citet{Willott1999} and modified by \citet{ShabalaGodfrey2013} who, by accounting for radiative losses, delivered a more correct relation. Therefore, the formula adopted by us is given as
\begin{equation}
\label{eq:Pj_Shabala}
\begin{aligned}
P_j\,{\rm [erg\,s^{-1}]} = & 1.5 \times 10^{43} \Bigg[ \bigg( \frac{\nu\,[\rm MHz]}{151}\bigg)^{\alpha} \frac{L_{1.4}\,{\rm [W\,Hz^{-1}]}}{10^{27}} \Bigg]^{0.8} \\
 & \times (1+z) \, (D\,{\rm [kpc]})^{0.58},
\end{aligned}
\end{equation}
where $L_{1.4}$ is the monochromatic lobe radio luminosity at $\nu = 1.4\,\rm GHz$, $\alpha = 0.8$ and $D$ is the source size. 

The conversion from radio luminosity to jet power derived by \citet{ShabalaGodfrey2013} is defined for FR\,II AGNs. These objects constitute 11\% of our sample and limiting to those most powerful radio galaxies would give us information about only a fraction of jetted AGNs, clearly biasing our understanding of the jet production mechanisms in various classes of AGNs. The only way to avoid this confusion is to obtain jet powers for all radio detected sources in our sample by establishing their upper limits. Hence, three groups can be identified: i) FR\,II type AGNs; ii) objects with lobed but not FR\,II type radio morphologies; iii) sources with radio detections but without visible double lobes. 

\begin{figure*}[ht!]
\begin{center}
\includegraphics[width=0.7\textwidth]{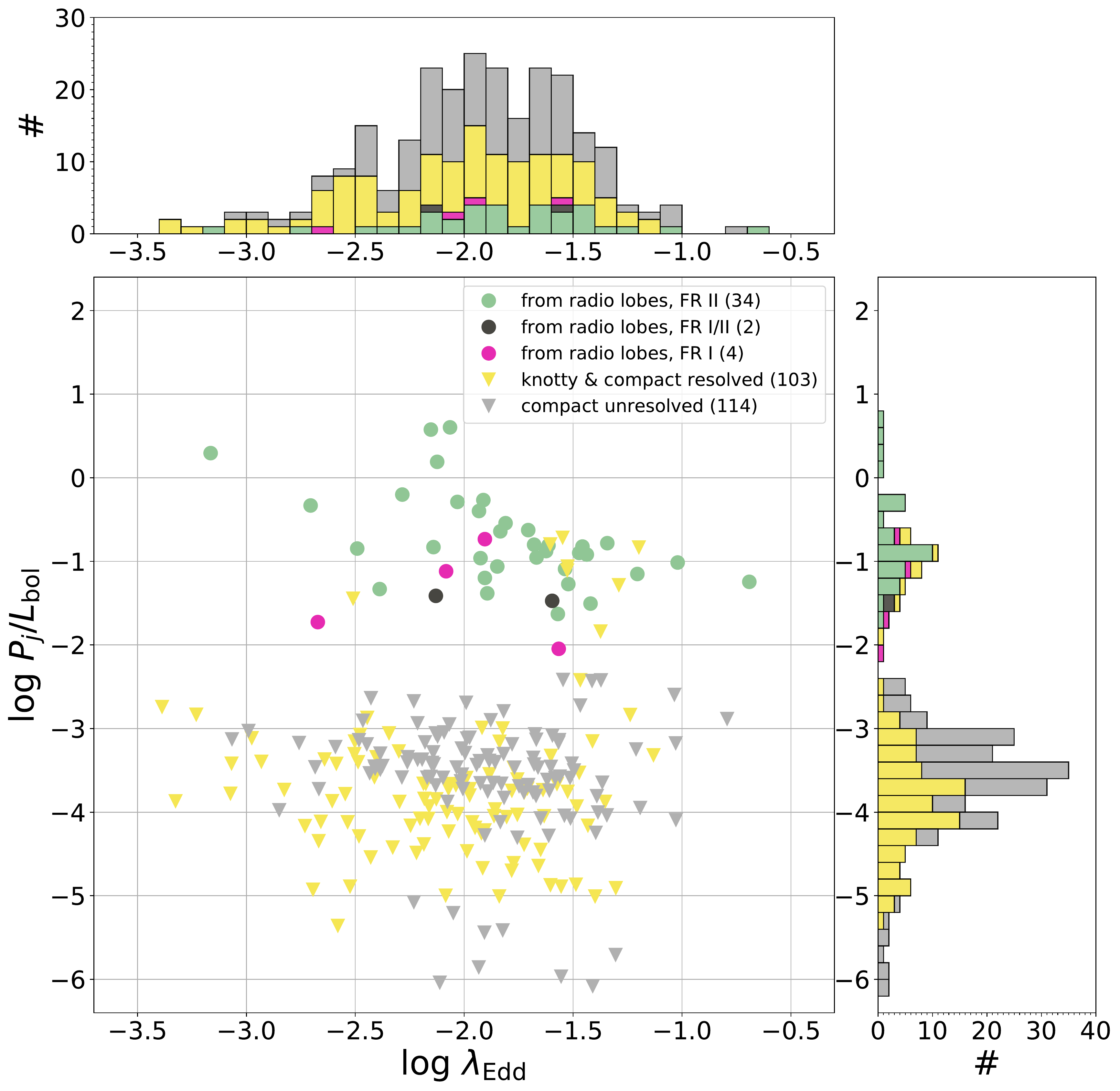}
\caption{The dependence of the $P_j/L_{\rm bol}$ ratio on the Eddington ratio $\lambda_{\rm Edd}$. The sample of 257 radio detected AGNs has been divided on the basis of how accurate the calculation of jet powers is, i.e. whether the radio emission from lobes can be extracted. The two main groups correspond to: lobed sources which are represented as filled colorful circles (40 objects altogether); non-lobed radio sources which are shown as filled colorful triangles (217 objects). Different colors for the group of lobed sources coincide with the Fanaroff-Riley classification as: green for FR\,II; dark grey for FR\,I/II; pink for FR\,I. Non-lobed radio objects are divided into knotty \& compact resolved and unresolved ones with yellow and grey colors, respectively.}
\label{fig_4_PjLbol_Eddratio}
\end{center}
\end{figure*}

A substantial difference appears in our calculations of lobed and non-lobed sources, as for the latter ones we use their total instead of lobe radio luminosity which was originally introduced in the Equation~(\ref{eq:Pj_Shabala}) and such a quantity can be estimated only for lobed AGNs. Regarding the extraction of lobe radio luminosities, we ensured (when possible) that it does not include any excess radio emission, such as the set of wings in X-shaped sources farther from the core, the outer pair of lobes in DDRGs, or any other radio emission regions which, based on their location on radio maps, do not belong to the radio lobes. In general, the accuracy of the jet power estimation decreases in each of the groups mentioned in the previous paragraph as we have less information about the exact radio characteristics (morphology, size and flux) of a given source. All jet power calculations for objects in our sample, scaled by bolometric luminosities, are shown in Fig.~\ref{fig_4_PjLbol_Eddratio}. 

Even though the results presented in Fig.~\ref{fig_4_PjLbol_Eddratio} rely mostly on upper limit estimations of $P_j$ (in 223 out of 257 objects), the bimodal distribution of $P_j/L_{\rm bol}$ is evident, noticeably separating objects powered by relativistic jets from sources in which the radio emission may be dominated by star formation, shocks generated by winds from accretion disks and their coronas, or by low-power jets \citep[and refs. therein]{Panessa2019}. No strong dependence between the Eddington ratio and scaled jet power is found in our sample. \\

\section{Host galaxies} 
\label{sec_HOST}

\begin{table*}[ht!]
\begin{center}
\caption{Host galaxy morphology of \textit{Swift}/BAT AGNs of different radio classes. The detailed description of host galaxies is given in Section~\ref{sec_HOST}.}
\label{tbl_2_radio_host}
\hskip-1.0cm
\begin{threeparttable}
\begin{tabular}{cccccccc} \hline
\multirow{2}{*}{Radio Class} & \multicolumn{6}{c}{Host Galaxy Morphology} & \multirow{2}{*}{Total} \\ \cline{2-7}
  & Elliptical & Lenticular & Spiral & Distorted & Merger & Unknown \\ \hline \hline 
RL         & 17 & 1 &     &  7           &    & 19 &  44  \\ 
RI\,\&\,RQ & 45 & 5 & 126 & 40$^{\star}$ &  7 & 47 & 270  \\ \hline \hline 
Total      & 62 & 6 & 126 & 47           &  7 & 66 & 314  \\ \hline
\end{tabular} 
\hskip2.2cm
\textbf{Note.} The source attributed to the irregular group is indicated by a star.
\end{threeparttable}
\end{center}
\end{table*}

Having well defined radio properties for most of the objects in our sample we checked what the characteristics of their optical counterparts were, specifically their host galaxy morphologies. In order to accomplish that we decided to use data from HST, SDSS, Pan-STARRS, and ESO archives finding that such information is available for 248 out of 314 AGNs, for which we were able to determine the host galaxy type spanning the
entire range of redshifts of our full sample\footnote{In fact all of our objects were found in the given surveys but not all of them had good enough data to establish their host morphology resulting in the exclusion of 66 objects.}. A detailed description of the data and the procedure used to establish the morphologies of host galaxies is given in Appendix~\ref{appendix_HOST} and below we describe the most important results. 

Based on the appearance of the host galaxy in the optical image we distinguish five morphological groups: \textit{elliptical}; \textit{lenticular}; \textit{spiral}; \textit{distorted}; and \textit{merger}\footnote{Additionally we identify one more group, \textit{irregular}, consisting of only one source, 2MASX J23444387-4243124. We include this object in the distorted group.}. The last two groups refer to galaxies in which we were not able to attribute any of the Hubble morphological types. We call a galaxy distorted when its morphology is disarranged, most probably resulting from galaxy interactions, but with only one nucleus. Galaxies in which two nuclei are present are classified as mergers. Some signs of galactic interactions (like tails, bridges, or small distortions) are also seen in both, ellipticals and spirals, constituting $\sim 11\%$  of each of these types. 

In Table~\ref{tbl_2_radio_host} we show the above described host galaxy morphological classification with regard to AGNs with and without powerful jets, represented by RL and RI\,\&\,RQ sources, respectively. Keeping in mind that we have information for $57\%$ and $83\%$ sources in RL and RI\,\&\,RQ groups, correspondingly, we see that the majority of radio-loud objects are found to reside in ellipticals while none are found in spirals which is not the case for radio-quieter AGNs, where both ellipticals and spirals are present with a prevalence of the latter. Such an observed lack of spiral morphologies in AGNs with extended radio structures is well-documented \citep[e.g.][]{WilsonColbert1995,McLure2004,Best2005,Madrid2006,WolfSheinis2008} albeit some studies reveal several radio lobed AGNs showing clear disk and/or spiral morphologies in optical images \citep[see Table 4 in][and refs. therein]{Tadhunter2016}. The fraction of objects with disturbed morphologies (distorted and merger) in RL and RI\,\&\,RQ AGNs is not very different being $28\%$ and $21\%$, respectively. 

We note that our results on host galaxies of \textit{Swift}/BAT AGNs and their relation to radio properties should be treated as the first step towards more extensive research which we plan to proceed with in the future.

\section{Theoretical implications} 
\label{sec_DISCUSSION}

While a consensus is almost reached that relativistic jets in RL AGNs are produced involving the Blandford-Znajek mechanism \citep[BZ; see review by][and refs. therein]{Blandford2019}, we are still lacking answers for such basic  questions as:
\begin{itemize}
\item[(1)] what is the dominant driver of the very large diversity of jet production efficiencies indicated by radio observations?
\item[(2)] is there any threshold required for production of jets in RL AGNs?
\item[(3)] are the weak jets observed in some RQ AGNs produced by the same mechanism as RL AGNs?
\item[(4)] why are powerful jets preferentially produced in AGNs hosted by elliptical galaxies?
\end{itemize}

Noting that the rate of energy extraction from rotating BH  by the BZ mechanism is $P_{\rm BZ} \propto  a^2 \Phi_{\rm BH}^2$, where $0<a<1$  is the dimensionless BH spin and $\Phi_{\rm BH}$ is the magnetic flux confined on the BH by the accretion flow, one may investigate the two following 'edge' scenarios to try to explain the diversity of jet production efficiency: the ``spin paradigm'' -- according to which the diversity of the energy extraction rate is driven by the spread in BH spins; and the ``magnetic flux paradigm'' -- where the  diversity is determined by the amount of magnetic flux threading the BH.

\subsection{Spin Paradigm?}

Albeit very popular \citep[][ and refs. therein]{WilsonColbert1995,Sikora2007,Fanidakis2011,Schulze2017}, the ``spin paradigm'' is seriously challenged by the fact that in order to explain the spread of the jet production efficiency by at least 3 orders of magnitude, the average BH spin in RQ AGNs should be smaller than $0.03$, while that estimated using the 'Soltan-type argument' is predicted to be $\sim0.6$ \citep{Soltan1982,ChokshiTurner1992,SmallBlandford1992,Elvis2002,YuTremaine2002,Lacy2015}. Similarly large BH spins were found by simulations of the cosmological evolution of BHs \citep{Volonteri2007,Volonteri2013}. Hence in order to reconcile the spin paradigm with radio-quiet AGNs having spins as large as $\sim 0.6$, the dependence of the jet production efficiency on the spin for larger values should be much stronger than quadratic. That possibility was recently claimed to be achievable by \citet{UnalLoeb2020} in the model, according to which magnetic fields threading the BH are anchored in the accretion disk. However, noting that magnetic tubes generated in the accretion disk carry on average zero magnetic flux, the jet is predicted to be produced in a flaring fashion \cite[see, e.g.][]{Yuan2019,Mahlmann2020} and the resulting time-averaged jet powers can be much smaller than deduced from obervations.

\subsection{Magnetic Flux Paradigm?}

The above might imply that the large diversity of the jet production efficiency is primarily determined by the amount of magnetic flux collected on the BH. However in order to convert the electromagnetic outflow generated by the BZ mechanism into narrow, relativistic jets, external confinement is required \citep{Beskin1998,Chiueh1998,LyubarskyEichler2001}. Such a confinement can be provided by magnetohydrodynamic (MHD) outflows from accretion disks and in the case of powerful jets is presumably associated with magnetically arrested disks \citep[MADs, e.g.][]{Narayan2003,Tchekhovskoy2011,McKinney2012}. Since MADs are formed only if the centrally accumulated magnetic flux exceeds the maximum amount which can be confined on the BH by the accretion flow, the division of AGNs into RL and RI/RQ ones is likely to correspond to the division of AGNs with and without MADs, or, equivalently, that the formation of the MAD provides a sort of threshold for launching powerful relativistic jets.

\subsection{Jet Powers in AGNs with MADs}

In order to reconcile such a 'MAD-nonMAD' bimodality with our calculated distribution of the jet production efficiency tracer $P_j/L_{\rm bol}$ (see Fig.~\ref{fig_4_PjLbol_Eddratio}), we need to explain what is the cause of the 2 dex span of this ratio for the RL AGNs assuming they all have MADs. Such a large spread can be an artifact resulting from the calculation of jet powers using their statistical correlation with their radio luminosities \citep{Willott1999}. While adequate in a statistical sense, the conversion formula based on such a correlation may give very large errors for individual sources. Those errors can be associated with the possible spread of such parameters as matter content, minimum electron energy \citep{Willott1999}, cooling effects \citep{ShabalaGodfrey2013}, and density of the environment into which the lobes are inflated \citep{HardcastleKrause2013}. But an intrinsic spread of the jet powers in MAD-AGNs is also expected,  contributed to by spreads in the BH spin and in the efficiency of the jet collimation by the MHD outflows powered by MADs with different sizes. 

It is encouraging to see a similar distribution of $P_j/L_{\rm bol}$ for our RL AGN sample (Fig.~\ref{fig_4_PjLbol_Eddratio}) and for those found by \citet{vanVelzenFalcke2013} and \citet{Inoue2017} for radio-loud quasars. They all peak at $P_j/L_{\rm bol} \sim 0.1$ despite the AGNs included in these samples covering very different accretion rates \citep[see also][were four various samples of RL AGNs were studied]{Rusinek2017}. Such an average value of the jet production efficiency is smaller by a factor $\sim 100$ than the maximum predicted for MAD-AGNs by numerical simulations \citep[e.g.][]{Tchekhovskoy2011}. However it should be noted that the fraction of maximal magnetic flux confined on the BH by the accretion flow depends on the geometrical thickness of the accretion flow \citep[see][]{Avara2016} and that $P_j \sim \dot M c^2$ (i.e. $P_j \sim  10 L_{\rm bol}$ for $\epsilon_d \sim 0.1$) is achievable only for geometrically thick accretion flows which are not representative for our AGNs nor for quasars.

\subsection{Jets in RQ AGNs}

As we can see in Table~\ref{tbl_1_radio_morph}, whereas most of the RL AGNs in our sample are extended (40/44), most of RI AGNs are compact (15/20). Hence there are no doubts that the deficiency of RI AGNs with extended structures is real. Such a deficiency does not mean that there is some threshold for operation of the BZ mechanism, but may simply reflect the very inefficient collimation in AGNs without the help of MHD winds from MADs. Then, the badly collimated BZ outflows would be significantly entrained by winds from stars enclosed within the outflow volume, slowed down and shocked, and then most their kinetic energy would be expected to be converted to the plasma heat, rather than used to accelerate relativistic electrons producing synchrotron radiation. Such radio sources, together with accretion disk coronas, presumably represent the compact radio sources observed in RQ and RI AGNs, with others being associated with SFRs and accretion disk winds/jets \citep[see review by][and refs. therein]{Panessa2019}. Some insights into the nature of compact radio sources of RQ \textit{Swift}/BAT AGNs are provided by \citet{Smith2016,Smith2020}.

\subsection{Building-up the MAD}

One can envision the following scenarios to form a MAD: by advection of the magnetic flux by accretion flows; by accumulation of sufficiently large magnetic flux in a galactic core prior to triggering the AGN phase; by building up the MAD by the so called ``Cosmic Battery''. 

Advection of poloidal magnetic fields by accretion disks was predicted to be very efficient by \citet{Bisnovatyi-KoganRuzmaikin1976}. Such a possibility was questioned by \citet{Lubow1994} who pointed out that due to the diffusion of magnetic fields in turbulent plasma such advection cannot proceed in geometrically thin disks. As later studies showed, poloidal magnetic fields can be advected by the surface layers of accretion disks \citep{Bisnovatyi-KoganLovelace2007,RothsteinLovelace2008,GuiletOgilvie2012,GuiletOgilvie2013,ZhuStone2018,CaoLai2019}. However, the efficiency of such advection can be limited at distances at which models of thin accretion disks predict their fragmentation due to gravitational instabilities \citep[e.g.][]{Hure1994,Goodman2003} and ambipolar diffusion in the outer, only partially ionized portions of the accretion flow \citep[e.g.][]{Begelman1995}. Finally, the advected poloidal magnetic field can be multi-polar and therefore the formation of a uni-polar magnetosphere over the BH and in the innermost portions of accretion disk -- the basic attributes of the MAD -- may require more time than the typical lifetime of the AGN, $t_{\rm AGN}$. Then, only AGNs which are at the age $t_{\rm AGN} > \Delta t_{\rm MAD}$ can produce powerful jets,  where $\Delta t_{\rm MAD}$ is the time it takes to form the MAD. This possibility seems to be supported by noting that the typical lifetimes of FR\,II sources are $\sim 3 \times 10^7$ years \citep[e.g.][]{Bird2008}, while the lifetimes of RQ AGNs are presumably much shorter \citep[e.g.][]{Schawinski2015,Schmidt2018,Khrykin2019}. However, it should be noted that the prevalence of MADs in ellipticals over disk galaxies, despite their AGNs having similar lifetimes, can be explained by a larger amount of advected magnetic flux per unit mass in the former. 

Regarding the second scenario, \citet{Sikora2013} proposed that the central accumulation of magnetic flux occurs during a hot accretion phase, prior to a cold, higher accretion rate 'event' representing the AGN phase. As suggested by \citet{SikoraBegelman2013} such an event might be triggered by the merger of a giant elliptical galaxy with a disk galaxy. 

Finally, MADs can be formed locally via operation of the Cosmic Battery \citep[CB,][]{ContopoulosKazanas1998,KoutsantoniouContopoulos2014,Contopoulos2018}. The model is based on the Poynting-Robertson radiation drag effect which generates a non-zero component of the toroidal electric field in the innermost regions of the accretion flow,  which in turn gives rise to the growth of poloidal magnetic field loops. Assuming that the outer parts of the loops diffuse outwards, the inner parts must then accumulate onto the BH. The possibility of the formation of a MAD by CB was numerically confirmed by GRMHD (general-relativistic magnetohydrodynamic) simulations \citep{Contopoulos2018}, however, so far only for non-rotating BHs and optically thin accretion flows. Unfortunately, the analytically estimated time scales for building up the MAD for  BHs with masses larger than $10^9 M_{\odot}$ exceeds the Hubble time \citep{Contopoulos2018}. But, noting that the Poynting-Robertson effect can be much stronger in the case of counter-rotating disks, one cannot exclude the formations of MADs in such AGNs within their lifetime. Combining this possibility with predictions that counter-rotating configurations can only be formed following mergers involving giant ellipticals with gas-rich spirals \citep{Garofalo2020}, one can explain why radio-loud AGNs are extremely rarely hosted by spiral galaxies (see Section~\ref{sec_HOST}). Furthermore, since the fraction of counter-rotating disks formed in such mergers is  expected to be $<50\%$, one can also explain why radio-quiet AGNs can be found in both ellipticals and spirals.\footnote{The idea of having RL AGNs associated with BHs and accretion disks rotating in opposite directions to each other is not new.  It was proposed by \citet{Garofalo2010}, based on the works by \citet{Reynolds2006} and \citet{Garofalo2009a,Garofalo2009b} who argued that jets produced by the BZ mechanism are more powerful in systems with retrograde disks than in systems with prograde disks. They suggested that this can explain the radio bimodality of AGNs. However, GRMHD simulations showed that the largest jet powers are achievable not in retrograde configurations, but in prograde configurations \citep{TchekhovskoyMcKinney2012,Tchekhovskoy2012}. This might suggest that, opposite to what was proposed by \citeauthor{Garofalo2010}, RL AGNs should be associated with prograde disks, and, therefore, the fraction of AGNs with retrograde disks should be much larger than the fraction of AGNs with prograde disks. However, the difference of maximal jet powers (those produced assuming the MAD models) produced by AGNs with retro- and prograde disks is much too small (only by a factor $\sim 3$) to explain the observed jet production diversity and, therefore, cannot be responsible for the radio loudness distribution with RL and RQ peaks observed to be separated by a factor $\sim 500$.}

\newpage
\section{Summary}
\label{sec_SUMMARY}

In this paper we investigate in detail the radio properties of massive AGNs ($M_{\rm BH} \geq 10^{8.5} M_{\odot}$) studied by \citet{Gupta2020} selected from the \textit{Swift}/BAT catalog \citep{Ricci2017}. Such a sample is excellent for justifying the radio bimodality, claimed by some but questioned by others, and the resulting diversity of jet production efficiency for several reasons. First of all, by selecting via matching the \textit{Swift}/BAT sources with galaxies, one is avoiding biases associated with radio and optical selection effects; secondly -- most objects in our sample are located at redshifts $z< 0.2$, which allows for the study of their radio and optical morphologies; thirdly -- by excluding high accretion rates, we allow the calculation of masses of their host galaxies and, consequently, BH masses using NIR luminosities; fourthly -- having the radiative output  of our objects be dominated by the MIR, given by WISE, and hard X-rays, given by BAT, allowed for reliable estimations of bolometric luminosities; and finally -- by excluding Compton-thick AGNs in our sample, we were able to verify the isotropy of some radiative features by comparing them to Type 1 and Type 2 AGNs \citep[see][]{Gupta2020}. Obviously, due to the very low sensitivity of \textit{Swift}/BAT the size of our sample is very much limited and therefore the presented results must be treated with some caution, but at the same time they show incredible potential for future use of AGNs selected by e\_ROSITA survey.

Our main results and their interpretation can be summarised as follows:

$\bullet$ the distribution of radio loudness in our studied \textit{Swift}/BAT AGN sample studied is bimodal, with the RL AGNs being on average 500 times radio-louder than RQ AGNs;

$\bullet$ assuming the same relation between radio luminosity and jet power for the entire sample, the distribution of jet production efficiency and of its upper limits were determined;

$\bullet$ a deficiency of jets with intermediate jet production efficiency implies the existence of threshold conditions for the production of powerful jets;

$\bullet$ our premise is that such conditions can be associated with formation of MADs, and that only those AGNs  which live longer than the time required to build up the MAD can become RL;

$\bullet$ the extremely rare cases of having RL AGNs hosted  by spiral galaxies, and  having RQ AGNs hosted by both spiral and elliptical galaxies seems to favor the scenario, according to which the MAD is built up by the Cosmic Battery and can be accomplished  within the AGN lifetime only in AGNs with accretion disks rotating in the opposite direction to the BH.

\acknowledgments

We thank David Abarca for useful discussions and editorial assistance. The research leading to these results has received funding from the Polish National Science Centre grant 2016/21/B/ST9/01620 and NAWA (Polish National Agency for Academic Exchange) grant PPN/IWA/2018/1/00100. 

\textit{Facilities:} Based on observations made with the NASA/ESA Hubble Space Telescope, and obtained from the Hubble Legacy Archive, which is a collaboration between the Space Telescope Science Institute (STScI/NASA), the Space Telescope European Coordinating Facility (ST-ECF/ESA) and the Canadian Astronomy Data Centre (CADC/NRC/CSA).

Funding for the Sloan Digital Sky Survey IV has been provided by the Alfred P. Sloan Foundation, the U.S. Department of Energy Office of Science, and the Participating Institutions. SDSS-IV acknowledges support and resources from the Center for High-Performance Computing at the University of Utah. The SDSS web site is www.sdss.org.

SDSS-IV is managed by the Astrophysical Research Consortium for the Participating Institutions of the SDSS Collaboration including the Brazilian Participation Group, the Carnegie Institution for Science, Carnegie Mellon University, the Chilean Participation Group, the French Participation Group, Harvard-Smithsonian Center for Astrophysics, Instituto de Astrof\'isica de Canarias, The Johns Hopkins University, Kavli Institute for the Physics and Mathematics of the Universe (IPMU) / University of Tokyo, the Korean Participation Group, Lawrence Berkeley National Laboratory, Leibniz Institut f\"ur Astrophysik Potsdam (AIP), Max-Planck-Institut f\"ur Astronomie (MPIA Heidelberg), Max-Planck-Institut f\"ur Astrophysik (MPA Garching), Max-Planck-Institut f\"ur Extraterrestrische Physik (MPE), National Astronomical Observatories of China, New Mexico State University, New York University, University of Notre Dame, Observat\'ario Nacional / MCTI, The Ohio State University, Pennsylvania State University, Shanghai Astronomical Observatory, United Kingdom Participation Group, Universidad Nacional Aut\'onoma de M\'exico, University of Arizona, University of Colorado Boulder, University of Oxford, University of Portsmouth, University of Utah, University of Virginia, University of Washington, University of Wisconsin, Vanderbilt University, and Yale University.

The Pan-STARRS1 Surveys (PS1) and the PS1 public science archive have been made possible through contributions by the Institute for Astronomy, the University of Hawaii, the Pan-STARRS Project Office, the Max-Planck Society and its participating institutes, the Max Planck Institute for Astronomy, Heidelberg and the Max Planck Institute for Extraterrestrial Physics, Garching, The Johns Hopkins University, Durham University, the University of Edinburgh, the Queen's University Belfast, the Harvard-Smithsonian Center for Astrophysics, the Las Cumbres Observatory Global Telescope Network Incorporated, the National Central University of Taiwan, the Space Telescope Science Institute, the National Aeronautics and Space Administration under Grant No. NNX08AR22G issued through the Planetary Science Division of the NASA Science Mission Directorate, the National Science Foundation Grant No. AST-1238877, the University of Maryland, Eotvos Lorand University (ELTE), the Los Alamos National Laboratory, and the Gordon and Betty Moore Foundation.

\newpage
\appendix
\section{The data}
\label{appendix_DATA}

\subsection{Radio data} 
\label{subsec_appendix_DATA_radio}

Given that the sources in \citet{Gupta2018} are from the northern as well as southern hemisphere, radio data was collected from two catalogs: National Radio Astronomy Observatory (NRAO) Very Large Array (VLA) Sky Survey \citep[NVSS,][]{Condon1998}; and Sydney University Molonglo Sky Survey \citep[SUMSS,][]{Bock1999, Mauch2003}. Both are characterized by similar sensitivity ($\sim$2.5\,mJy) and resolution (45\,arcsec FWHM for NVSS and 45$\times$45\,cosec$|\delta|$\,arcsec$^2$ for SUMSS). NVSS, a 1.4\,GHz continuum survey, covers the northern sky from $-40$ deg declination while SUMSS, a wide-field radio imaging survey conducted at 843\,MHz, covers the southern sky from $-30$ deg declination, so together they map the whole sky. In addition to that, we decided to include one more catalog -- Faint Images of the Radio Sky at Twenty-cm \citep[FIRST,][]{Becker1995}. This 1.4\,GHz sky survey is distinguished by its high resolution (5.4\,arcsec) and its sensitivity down to 1\,mJy radio flux. It covers a piece of the sky surveyed by NVSS, therefore having detections in both of these catalogs helps to determine whether the source is compact or extended, but also to examine the accurate sizes of compact sources (see Appendix~\ref{appendix_SIZE} for more details).  

For NVSS and SUMSS data a search within a matching radius of 3 arcmin was conducted. For those sources, where a single association was found, its exact location was checked -- whether the radio match is located within 30 arcsec from the optical center and if so, this match was assigned to the object. The same procedure was adopted for FIRST data with the only difference being the internal matching radius of 5 instead of 30 arcsec. All the sources having more than one radio association within a matching radius of 3 arcmin were checked by eye. This part was done through visual inspection of radio maps with size of 0.45\,deg $\times$ 0.45\,deg extracted from NVSS\footnote{http://www.cv.nrao.edu/nvss/postage.shtml}, SUMSS\footnote{http://www.astrop.physics.usyd.edu.au/cgi-bin/postage.pl} and FIRST\footnote{https://third.ucllnl.org/cgi-bin/firstcutout} by using the NRAO AIPS (Astronomical Image Processing System\footnote{http://www.aips.nrao.edu/}) package. In order to avoid false associations of radio matches we downloaded the images from DSS1 and DSS2 which are digitized versions of several photographic astronomical surveys, in addition to radio maps, which can be found at the ESO archive\footnote{http://archive.eso.org/dss/dss}. Comparison of radio and optical sources on maps from both domains, together with NED (NASA/IPAC Extragalactic Database\footnote{https://ned.ipac.caltech.edu/}), enabled us to distinguish incorrect matches. As some sources turned out to have extremely extended, i.e. beyond 3 arcmin, radio structures we were gradually increasing the radio search by 1 arcmin as long as the association for the whole structure was found. Once the whole radio emitting region was identified, the radio flux of each of the components was summed up and assigned to the given source. For sources lacking radio detections we assigned them upper limits corresponding to the value of the sensitivity of the survey which contains the source in its footprint. While the sensitivity of NVSS and SUMSS is the same, for objects located in the area covered by both NVSS and FIRST, we decided on FIRST upper limits, as its sensitivity is lower than that of NVSS.

Since the radio catalogs we used have been conducted at two different frequencies, the radio fluxes at 843\,MHz were recalibrated to 1.4\,GHz using a radio spectral index of $\alpha_r = 0.8$ (with the convention of $F_{\nu} \propto \nu^{-\alpha}$), so that the rest of our calculations are consistent.

\subsection{Mid-infrared data} 
\label{subsec_appendix_DATA_MIR}

The mid-infrared measurements were taken from the AllWISE Data Release \citep{Cutri2013} which, by combining data from cryogenic Wide-field Infrared Survey Explorer \citep[WISE,][]{ Wright2010} and post-cryogenic NEOWISE \citep[``near-Earth object + WISE'',][]{Mainzer2011} survey phases, resulted in a comprehensive view of the mid-infrared sky. Out of four available bands (at 3.4, 4.6, 12 and 22\,$\upmu$m, corresponding to the W1, W2, W3 and W4 bands, respectively), we decided to make use of the W3 band, which was driven by the fact that at this specific wavelength the dusty torus is transparent enough to observe radiation coming directly from the central region of an AGN. At the shorter wavelengths of the W1 and W2 bands the dusty, circumnuclear tori are optically thick and radiate anisotropically \citep{Honig2011,Netzer2015}. At the longer wavelengths of the W4 band, the dusty torus becomes even more transparent, but its measurements are affected by much larger errors than in W3 because: (1) the W4 band traces the warm dust continuum at 22\,$\upmu$m, and can be contributed to by starbursts \citep{Ichikawa2019}; (2) the sensitivity in the W4 band is much lower than in the W3 band and its signal-to-noise ratio is the lowest of all the WISE channels, resulting in a lower detected fraction for the objects in our sample; and (3) the resolution of W4 images is worse than in other bands (e.g. 12 arcsec in W4 vs. 6.5 arcsec in W3), so for some of our sources (point-like in optical, with close neighbours) W4 images could be blended.

Searching for counterparts was conducted within a radius of 5 arcsec, which allowed for the avoidance of false matches as the angular resolution of WISE in W3 band is 6.5 arcsec. 

The conversion from W3 magnitude, $m_{\rm W3}$, to the monochromatic flux, $F_{\nu_{\rm W3}}$, was done following the formula provided by \citet{Wright2010}, given as $F_{\nu_{\rm W3}} = 30.922 \times 10^{(-m_{\rm W3}/2.5)}$.

\section{Size estimation for compact AGNs}
\label{appendix_SIZE}

The radio catalogs we used differ significantly not only in their angular resolutions, and consequently in the accuracy of their fitted deconvoled angular sizes, but also in the data provided for unresolved objects. In NVSS, the angular size is estimated for each object, either as an exact measurement or an upper limit. However, in the case of FIRST and SUMSS, for unresolved sources a size of $0.0$ arsec is assigned. Therefore, the smallest resolved size listed in each of these catalogs, i.e. $0.01$ and $17.3$ arcsec for FIRST and SUMSS, respectively, was taken as an upper limit for unresolved objects detected in these catalogs. Additionally, as was already mentioned in Section~\ref{subsec_RADIO_radio_loudness}, within a group of 57 compact sources for which data in NVSS and FIRST was available: (1) the angular sizes from NVSS are on average 15 times bigger than those taken from FIRST; (2) all but 4 objects are resolved in FIRST (52/57) and only 8 are resolved in NVSS (8/57). 

Taking into account the above, we decided to use the following procedure to obtain sizes for compact sources:
\begin{itemize}
\item[--] for sources having both NVSS and SUMSS measurements, the NVSS size was adopted (8 sources);
\item[--] for sources having both NVSS and FIRST measurements, the FIRST size estimate was adopted (57 sources);
\item[--] there are 13 sources with detections only in FIRST, 96 with detections only in NVSS, and 31 with detections only in SUMSS. 
\end{itemize}

\section{Optical data for host galaxy classification}
\label{appendix_HOST}

Optical images for sources in our sample were collected from the following resources: Hubble Space Telescope (HST) through the Hubble Legacy Archive\footnote{http://hla.stsci.edu/hlaview.html}; Sloan Digital Sky Survey (SDSS) Data Release 14 \citep[DR14,][]{Abolfathi2018} through ImgCutout Web Service\footnote{http://skyservice.pha.jhu.edu/dr14/ImgCutout/ImgCutout.asmx}; Panoramic Survey Telescope and Rapid Response System \citep[Pan-STARRS,][]{Chambers2016} through Image Cutout Server\footnote{https://ps1images.stsci.edu/cgi-bin/ps1cutouts}. For galaxies that did not have images available in any of those services, we examined (mostly near-infrared) images retrieved from the ESO Archive Science Portal\footnote{http://archive.eso.org/scienceportal/}.

Similarly to radio data, here we also make use of data differing in angular resolution, optical filter, and sensitivity to extended emission, deciding that the most and the least advantageous information is retrieved from HST and Pan-STARRS data, respectively. This translates to the 'importance' of the data in the order of: HST as the most reliable; SDSS; and Pan-STARRS as the most uncertain. Objects with data taken from ESO archives are excluded from this 'sequence' as: (1) they are found in only one resource; (2) their classification is based on mainly near-infrared, not optical, images. 

\renewcommand{\thetable}{\Alph{section}.1}

\begin{table*}[ht!]
\begin{center}
\caption{The qualitative confidence of host galaxy morphology classification for 248 AGNs from our sample.}
\label{tbl_appendix_1_host_reliability}
\hskip-1.0cm
\begin{threeparttable}
\begin{tabular}{ccccccc} \hline
\multirow{2}{*}{Reliability} & \multicolumn{5}{c}{Host Galaxy Morphology} & \multirow{2}{*}{Total}\\ \cline{2-6}
 & Elliptical & Lenticular & Spiral & Distorted & Merger & \\ \hline \hline
0     &  3 &    &  18 &  6           & 1 &  28 \\
1     & 41 & 2  &  94 & 29$^{\star}$ & 6 & 172 \\
2     &  4 &    &   4 &  5           &   &  13 \\
3     &  4 & 2  &   1 &  2           &   &   9 \\
4     & 10 & 2  &   9 &  5           &   &  26 \\ \hline \hline
Total & 62 & 6  & 126 & 47           & 7 & 248 \\ \hline
\end{tabular} 
\hskip2.0cm
\textbf{Note.} The source attributed to the irregular group is indicated by a star.
\end{threeparttable}
\end{center}
\end{table*}

Knowing about the origin of the data and based on how many details of the host galaxy can be determined, how accurate the optical images are and how many of them were available for a given source we introduced the reliability flag describing the qualitative confidence of our classification. Five groups correspond to the following:
\begin{itemize}
\item[0 --] when the optical images from all three resources were available and the morphology type is obvious and the same in all of them,
\item[1 --] when the optical images from one or two optical resources were available and the morphology type is obvious and the same in all of them,
\item[2 --] when the optical images from two or three resources were available and the morphology type differs between them being well defined in the image coming from the most credible resource,
\item[3 --] when the optical images from two or three resources were available and the morphology type differs between them being unreliable in the image coming from the most credible resource,
\item[4 --] when the optical images from one or two resources were available and the morphology type seen there is doubtful.
\end{itemize}
In general the reliability of our classification decreases in each of the above groups as it is based on less solid information. Sources with ESO data are assigned to groups 1 or 4 only.

The exact numbers of AGNs flagged as described above are presented in Table~\ref{tbl_appendix_1_host_reliability}. A small fraction of all the objects with defined host galaxy morphology belongs to last three groups corresponding to those with the least reliability (48 out of 248). The classification of most of the sources seem to be quite robust (200 out of 248), although  optical images from all three resources exhibiting the same morphological type were only available for a few of them (28 AGNs).

\section{The sample}
\label{appendix_SAMPLE}

Table~\ref{tbl_appendix_2_whole_sample}.1 lists the most important information obtained and discussed in this work for some of the sources in our sample of \textit{Swift}/BAT AGNs. The complete sample is available as supplementary
material online.

\renewcommand{\thetable}{\Alph{section}.1}

\movetabledown=2.7in
\begin{table*}[h]
\begin{rotatetable*}
\label{tbl_appendix_2_whole_sample}
\begin{center}
\caption{\textit{Swift}/BAT sample of 314 AGNs used in this study.}
\scriptsize
\begin{tabular}{lccccccrrcrrcccc} \hline
\multirow{2}{*}{SWIFT Name}  & \multirow{2}{*}{Counterpart Name} & \multirow{2}{*}{$z$} & \multirow{2}{*}{$\log M_{\rm BH}$} & \multirow{2}{*}{$\log L_{\rm bol}$} & \multirow{2}{*}{$\log \lambda_{\rm Edd}$} & Radio & \mc{\multirow{2}{*}{$F_{\rm 1.4}$}} & \mc{\multirow{2}{*}{$R$}} & Radio & \mc{\multirow{2}{*}{Size}} & \mc{\multirow{2}{*}{$\log(P_{\rm j}/L_{\rm bol})$}} & Radio & \multirow{2}{*}{FR$^d$} & Host & Host  \\ 
  &  &  &  &  &  & Flag$^a$ &  &  & Class$^b$ &  &  & Morphology$^c$ &  & Flag$^e$ & Morphology$^f$ \\
  &  &  & ($M_{\odot}$) & (erg\,s$^{-1}$) &  &  & \mc{(mJy)} &  &  & \mc{(kpc)} &  &  &  &  &  \\ \hline \hline
J1940.4-3015 & IGR J19405-3016 & 0.0525 & 8.72 & 45.08 & -1.75 & N & 9.6 & -0.94 & RQ & 28.96$^{\dagger}$ & $<$-3.69 & C &  & P & S\\ 
J1952.4+0237 & 3C 403 & 0.0584 & 8.90 & 45.15 & -1.85 & N & 6045.4 & 1.88 & RL & 167.91 \ \ & -1.06 & X & II & H & L\\ 
J1959.4+4044 & Cygnus A & 0.0558 & 9.29 & 45.33 & -2.06 & N & 1598189.0 & 4.09 & RL & 104.65 \ \ & 0.60 & D & II & H & E\\ 
J2001.0-1811 & 2MASX J20005575-1810274 & 0.0372 & 8.57 & 45.27 & -1.40 & U & 2.5 & -2.03 & RQ &  &  &  &  & H & L\\ 
J2018.4-5539 & PKS 2014-55 & 0.0607 & 8.86 & 45.05 & -1.91 & S & 1597.5 & 1.44 & RL & 1367.75 \ \ & -0.73 & T & I & E & E\\ 
J2030.2-7532 & IRAS 20247-7542 & 0.114 & 9.02 & 46.09 & -1.03 & S & 8.6 & -1.29 & RQ & 35.80$^{\dagger}$ & $<$-4.10 & C &  &  & \\ 
J2033.4+2147 & 4C +21.55 & 0.174 & 9.17 & 45.73 & -1.54 & N & 1940.1 & 1.82 & RL & 544.32 \ \ & -1.09 & T & II & P & D\\ 
J2040.2-5126 & ESO 234-IG 063 & 0.0541 & 8.60 & 45.01 & -1.68 & S & 28.2 & -0.38 & RQ & 18.21$^{\dagger}$ & $<$-3.35 & C &  & E & D\\ 
J2042.3+7507 & 4C +74.26 & 0.105 & 9.32 & 45.90 & -1.52 & N & 1894.0 & 1.16 & RL & 1140.84 \ \ & -1.27 & T & II &  & \\ 
J2044.2-1045 & Mrk 509 & 0.0344 & 8.72 & 45.19 & -1.63 & N & 18.6 & -1.15 & RQ & 15.42 \ \ & $<$-4.05 & C &  & H & S\\ 
J2052.0-5704 & IC 5063 & 0.0115 & 8.56 & 44.74 & -1.92 & S & 1448.5 & 0.23 & RI & 9.53 \ \ & $<$-2.99 & K &  & H & E\\ 
J2109.2+3531 & B2 2107+35A & 0.202 & 9.30 & 45.88 & -1.52 & N & 1434.4 & 1.69 & RL & 344.87 \ \ & $<$-1.06 & K &  &  & \\ 
J2114.4+8206 & 2MASX J21140128+8204483 & 0.0833 & 8.87 & 45.41 & -1.57 & N & 474.9 & 0.84 & RI & 308.21 \ \ & -2.05 & T & I & P & E\\ 
J2116.3+2512 & 2MASX J21161028+2517010 & 0.153 & 8.62 & 45.05 & -1.67 & U & 2.5 & -0.51 & RQ &  &  &  &  &  & \\ 
J2118.9+3336 & 2MASX J21192912+3332566 & 0.0509 & 8.69 & 44.33 & -2.46 & N & 3.7 & -0.63 & RQ & 138.17$^{\dagger}$ & $<$-2.91 & C &  & P & E\\ 
J2134.9-2729 & 2MASX J21344509-2725557 & 0.0667 & 8.74 & 45.00 & -1.84 & N & 7.8 & -0.74 & RQ & 118.52 \ \ & $<$-3.16 & C &  & P & S$_{\rm i}$\\ 
J2135.5-6222 & 1RXS J213623.1-622400 & 0.059 & 8.72 & 45.17 & -1.65 & S & 4.3 & -1.28 & RQ & 19.74$^{\dagger}$ & $<$-4.08 & C &  &  & \\ 
J2137.8-1433 & PKS 2135-14 & 0.2 & 9.19 & 45.95 & -1.34 & N & 3864.0 & 2.04 & RL & 365.62 \ \ & -0.78 & D & II &  & \\ 
J2145.5+1101 & RX J2145.5+1102 & 0.209 & 8.73 & 45.25 & -1.57 & F & 1.1 & -0.77 & RQ & 16.66 \ \ & $<$-3.67 & C &  &  & \\ 
J2150.2-1855 & 6dF J2149581-185924 & 0.158 & 8.68 & 45.26 & -1.53 & N & 528.4 & 1.64 & RL & 220.81 \ \ & $<$-1.11 & C &  &  & \\ 
J2157.2-6942 & PKS 2153-69 & 0.0283 & 8.81 & 44.21 & -2.71 & S & 29203.8 & 2.86 & RL & 37.42 \ \ & -0.33 & T & II & H & E\\ 
J2200.9+1032 & Mrk 520 & 0.0275 & 8.54 & 44.78 & -1.86 & F & 60.8 & -0.42 & RQ & 1.57 \ \ & $<$-3.97 & C &  & S & D\\ 
J2204.7+0337 & 2MASX J22041914+0333511 & 0.0611 & 8.89 & 45.58 & -1.41 & F & 8.8 & -1.35 & RQ & 0.01$^{\dagger}$ & $<$-6.08 & C &  & S & D\\ 
J2209.1-2747 & NGC 7214 & 0.0227 & 8.87 & 44.38 & -2.59 & N & 28.2 & -0.52 & RQ & 13.66 \ \ & $<$-3.42 & C &  & P & S\\ 
J2214.2-2557 & 2MASX J22140917-2557487 & 0.0519 & 8.70 & 44.36 & -2.45 & N & 3.1 & -0.72 & RQ & 60.34$^{\dagger}$ & $<$-3.19 & C &  & P & S\\ 
J2217.0+1413 & Mrk 304 & 0.0704 & 8.96 & 45.26 & -1.80 & U & 2.5 & -1.44 & RQ &  &  &  &  &  & \\ 
J2223.9-0207 & 3C 445 & 0.0601 & 8.54 & 45.43 & -1.21 & N & 5783.9 & 1.60 & RL & 660.07 \ \ & -1.15 & X & II &  & \\ 
J2226.8+3628 & MCG +06-49-019 & 0.0213 & 8.61 & 43.64 & -3.07 & N & 7.0 & -0.44 & RQ & 18.96$^{\dagger}$ & $<$-3.13 & C &  & S & S\\ 
J2234.8-2542 & ESO 533- G 050 & 0.0265 & 8.67 & 43.53 & -3.24 & U & 2.5 & -0.59 & RQ &  &  &  &  & P & S\\ 
J2235.9+3358 & NGC 7319 & 0.0227 & 8.63 & 44.28 & -2.44 & N & 53.0 & -0.15 & RQ & 34.71 \ \ & $<$-2.87 & C &  & H & S$_{\rm i}$\\ 
J2246.0+3941 & 3C 452 & 0.0811 & 8.79 & 44.96 & -1.93 & N & 10556.6 & 2.62 & RL & 316.76 \ \ & -0.40 & T & II & H & E\\ 
J2248.7-5109 & 2MASX J22484165-5109338 & 0.1 & 8.52 & 45.34 & -1.28 & U & 2.5 & -1.20 & RQ &  &  &  &  &  & \\ 
 \hline  
\end{tabular}
\end{center}
\vspace{-0.2cm}
\textbf{Notes.} This subset of the table demonstrates format and content. \\
$^a$ Origin of radio flux: N -- NVSS; S -- SUMSS; F -- FIRST; U -- undetected. \\
$^b$ RL -- radio-loud; RI -- radio-intermediate; RQ -- radio-quiet. \\
$^c$ X -- complex; T -- triple; D -- double; K -- knotty; C -- compact.  \\
$^d$ Our Fanaroff-Riley classification for lobed sources only.\\
$^e$ Origin of host morphology classification: H -- HST; S -- SDSS ; P -- Pan-STARRS; E -- ESO. \\
$^f$ E -- elliptical; L -- lenticular; S -- spiral; D -- distorted; Irr -- irregular; M -- merger. Subscript 'i' indicates interacting galaxies. \\
$^{\dagger}$ Compact unresolved sources for which their angular sizes equal to upper limits obtained from a given catalog (Appendix~\ref{appendix_SIZE}).
\end{rotatetable*}
\end{table*}

\newpage 

\bibliographystyle{aasjournal}
\bibliography{refs}

\end{document}